\documentclass[12pt]{iopart}
\usepackage{graphicx,color}
\usepackage{verbatim}
\usepackage{xspace}
\usepackage{amsfonts}
\newcommand{\targetwave}{$\psi_{\mathrm{G}}$\xspace}
\newcommand{\trialwave}{$\Psi_{\mathrm{t}}$\xspace} 
\newcommand{\ctrialwave}{$\Psi_{\mathrm{t}}'$\xspace} 
\newcommand{\ehmol}{$\left(e^{+}-\mathrm{H}_{2}\right)$\xspace}
\newcommand{\hmol}{$\mathrm{H}_{2}$\xspace}
\newcommand{\ps}{$\eta_{\mathrm{v}}$\xspace}
\newcommand{\psh}{$\hat{\eta}_{\mathrm{v}}$\xspace}

\newcommand{\cond}{$\kappa\left(A\right)$\xspace}

\newcommand{\dist}{$\Lambda\left(A\right)$\xspace}
\newcommand{\diste}{$\Lambda_{\mathrm{a}}\left({A}\right)$\xspace}
\newcommand{\deter}{$\mathrm{det}\left(A\right)$\xspace}
\definecolor{myRed}{rgb}{0.9,0,0}
\definecolor{myGreen}{rgb}{0,0.4,0}

\begin{document}
 
\title[Anomaly-free singularities in the Kohn method]{Anomaly-free singularities in the generalized Kohn variational method}

\author{J N Cooper$^1$, E A G Armour$^1$ and 
M Plummer$^2$}

\address{$^1$ School of Mathematical Sciences, University Park, Nottingham NG7 2RD, UK}
\address{$^2$ STFC Daresbury Laboratory, Daresbury, Warrington, Cheshire WA4 4AD, UK}
\ead{james.cooper@maths.nottingham.ac.uk}

\begin{abstract}
We have carried out an analysis of singularities in Kohn variational calculations for low energy \ehmol elastic scattering. Provided that a sufficiently accurate trial wavefunction is used, we argue that our implementation of the Kohn variational principle necessarily gives rise to singularities which are not spurious. We propose two approaches for optimizing a free parameter of the trial wavefunction in order to avoid anomalous behaviour in scattering phase shift calculations, the first of which is based on the existence of such singularities. The second approach is a more conventional optimization of the generalized Kohn method. Close agreement is observed between the results of the two optimization schemes; further, they give results which are seen to be effectively equivalent to those obtained with the complex Kohn method. The advantage of the first optimization scheme is that it does not require an explicit solution of the Kohn equations to be found. We give examples of anomalies which cannot be avoided using either optimization scheme but show that it is possible to avoid these anomalies by considering variations in the nonlinear parameters of the trial function.
\end{abstract}

\submitto{\JPA}
\maketitle

\section{Introduction}\label{s:introduction}

Despite the absence of an explicit minimization principle, variational methods have been used successfully in many problems of quantum scattering theory. Such calculations typically exploit a stationary principle in order to obtain an accurate description of scattering processes. The Kohn variational method \cite{Kohn1948} has been applied extensively to problems in electron-atom \cite{Nesbet1980} and electron-molecule \cite{MasseyRidley1956,Schneider1988} scattering, as well as to the scattering of positrons, $e^{+}$, by atoms \cite{VanReeth1995,VanReeth1996,VanReeth1999} and molecules \cite{Armour1990,Cooper2007,CooperArmourPlummer2008}. It has been widely documented, however, that matrix equations derived from the Kohn variational principle are inherently susceptible to spurious singularities. These singularities were discussed first by Schwartz \cite{Schwartz1961,Schwartz1961b} and have subsequently attracted considerable attention \cite{Nesbet1968,Brownstein1968,Shimamura1971,Takatsuka1979}. In the region of these singularities, results of Kohn calculations can be anomalous.

Although sharing characteristics similar to those exhibited by scattering resonances \cite{Shimamura1971}, Schwartz singularities are nonphysical and arise only because the trial wavefunction, used in Kohn calculations to represent scattering, is inexact \cite{Nesbet1968}. For projectiles of a given incident energy, anomalous results are confined to particular formulations of the trial wavefunction and can, in principle, be mitigated by a small change in boundary conditions or some other parameter. It has also been shown \cite{Schneider1988,McCurdy1987,Lucchese1989} that the use of a complex-valued trial wavefunction avoids anomalous behaviour except in exceptional circumstances. Alternative versions of the Kohn method have been developed in terms of a Feshbach projection operator formalism \cite{Feshbach1962} and have been found \cite{Chung1971} to give anomaly-free results. 

In this article we will discuss our investigations of Schwartz-type anomalies for generalized Kohn calculations involving the elastic scattering of positrons by molecular hydrogen, $\mathrm{H}_{2}$. We will find that our choice of trial wavefunction contains a free parameter that can be varied in such a way as to produce singularities which are legitimate in the context of the scattering theory and which do not give rise to anomalous results. Indeed, these singularities can be used to formulate an optimization scheme for choosing the free parameter so as to automatically avoid anomalous behaviour in calculations of the scattering phase shift. The novelty of determining the phase shift in this way is that an explicit solution of the linear system of Kohn equations is not required. We will also develop an alternative optimization and show that the two schemes give results in close agreement. Further, the results obtained will be seen to be in excellent agreement at all positron energies with those determined via the complex Kohn method.
 
We will give examples of anomalous behaviour which cannot be avoided with either optimization, and show that the same anomalies appear in our application of the complex Kohn method. We will discuss circumstances under which these anomalies might occur. We will show also that such results are nonphysical by considering small changes in the nonlinear parameters of the trial wavefunction.

Our investigations of singular behaviour have been carried out as part of a wider study on \ehmol scattering and annihilation using extremely flexible wavefunctions. Our ability to recognize clearly and analyze the anomalous behaviour is as good for this system as it would be for a simpler model system, with the advantage that our calculations can be used to provide meaningful and physically relevant results \cite{CooperArmourPlummer2008}. 

\section{Theory}

\subsection{The generalized Kohn variational method}\label{ss:kvm}

The Kohn variational method is used to calculate approximations to exact scattering wavefunctions. Determining an approximation, \trialwave, allows a variational estimate, \ps, of the scattering phase shift to be calculated, the error in which is of second order in the error of \trialwave from the exact scattering wavefunction, $\Psi$ \cite{CharltonHumberston2005}. The standard approach in Kohn calculations is to assume an overall form for \trialwave that depends linearly on a set of unknown parameters, optimal values for which are then determined by the application of a stationary principle.

In our investigations of anomalous behaviour in Kohn calculations for \ehmol scattering, we have studied the lowest partial wave of $\Sigma_{\mathrm{g}}^{+}$ symmetry. This partial wave has been shown \cite{Armour1990} to be the only significant contributor to scattering processes for incident positron energies below $2$ eV. The first significant inelastic channel is positronium formation which has a threshold at $8.63$ eV \cite{CharltonHumberston2005}. Although we will here consider positron energies higher than these thresholds, it is not our intention to provide a comprehensive physical treatment of the scattering problem taking higher partial waves and inelastic processes into account. The purpose of the present study is to give a correct and, as far as possible, anomaly-free treatment of the lowest partial wave. It is important to examine the single channel case as accurately as possible as a preliminary for more sophisticated calculations. By not taking into account additional channels, it is possible that anomalous behaviour could occur due to physical inaccuracies in the trial wavefunction at higher energies. However, we will demonstrate that all of the anomalies in our results ultimately can be attributed to purely numerical effects.

We have initially used a trial wavefunction having the same general form as described in our earlier calculations \cite{Cooper2007},

\begin{equation}\label{eq:trialwave}
\Psi_{\mathrm{t}} = \left(\bar{S} + a_{\mathrm{t}}\bar{C} + p_{0}\chi_{0}\right)\psi_{\mathrm{G}} + \sum_{i=1}^{M} p_{i}\chi_{i},
\end{equation}
where

\begin{equation}\label{eq:tautrans}
\left[ \begin{array}{c}
{\bar{S}}\\
{\bar{C}}
\end{array}\right] =
\left[ \begin{array}{cc}
\cos(\tau) & \sin(\tau)\\
-\sin(\tau) & \cos(\tau)\\
\end{array}\right]
\left[ \begin{array}{c}
{S}\\
{C}\\
\end{array}\right],
\end{equation}
for some phase parameter, $\tau\in\left[0,\pi\right)$, with

\begin{equation}\label{eq:SinOpenChannel}
S=\frac{N}{\lambda_{3} - 1}\sin\left[ c\left( \lambda_{3} - 1\right) \right],
\end{equation}
and

\begin{equation}\label{eq:CosOpenChannel}
 C = \frac{N}{\lambda_{3} - 1}\cos\left[c\left( \lambda_{3}-1\right) \right] \lbrace 1-\exp\left[-\gamma\left( \lambda_{3}-1\right)\right]\rbrace.
\end{equation}
As before \cite{Cooper2007}, we have carried out calculations using the fixed-nuclei approximation \cite{Temkin1967,Temkin1969}, taking the internuclear separation to be at the equilibrium value, $R=1.4$ au. We have labelled the electrons as particles $1$ and $2$, taking the positron to be particle $3$. The position vector, $\mathbf{r}_{j}$, of each lepton is described by the prolate spheroidal coordinates \cite{Flammer1957} $(\lambda_{j},\mu_{j},\phi_{j})$, $j\in\{1,2,3\}$. These coordinates are defined implicitly in terms of the Cartesian coordinates, $(x_{j},y_{j},z_{j})$, as

\begin{eqnarray}
x_{j} &=& \frac{1}{2}R\left[\left( \lambda_{j}^{2}-1\right) \left( 1-\mu_{j}^{2}\right)\right]^{\frac{1}{2}} \cos \left(\phi_{j}\right),\\
y_{j} &=& \frac{1}{2}R\left[\left( \lambda_{j}^{2}-1\right) \left( 1-\mu_{j}^{2}\right)\right]^{\frac{1}{2}} \sin \left(\phi_{j}\right),\\
z_{j} &=& \frac{1}{2}R\lambda_{j} \mu_{j}.
\end{eqnarray}
The functions $S$ and $C$ represent, respectively, the incident and scattered positrons asymptotically far from the target. The shielding parameter, $\gamma$, ensures the regularity of $C$ at the origin and is taken to have the value $\gamma=0.75$. The constant, $c$, is defined to be $c=kR/2$, $k$ being the magnitude of the positron momentum in atomic units. $N$ is a normalization constant and can here be regarded as arbitrary. The unknowns, $a_{\mathrm{t}}$ and $\{p_{0},\dots,p_{M}\}$, are constants to be determined. The inclusion of the parameter, $\tau$, in \trialwave is a generalization of the Kohn method due to Kato \cite{Kato1950,Kato1951}. This parameter is of only minor physical significance, playing the role of an additive phase factor in the part of the wavefunction representing the incident and scattered positrons asymptotically far from the target. However, at each value of $k$, the value of $\tau$ can be varied to avoid spurious singularities in the Kohn calculations. Away from the spurious singularities, for an accurate trial wavefunction we can expect the variation in the calculated values of \ps over $\tau$ to be small. In the original application of the Kohn method \cite{Kohn1948}, only wavefunctions corresponding to $\tau=0$ were considered.

The function, \targetwave, is an approximation to the ground state wavefunction of the unperturbed hydrogen molecule and is determined by the Rayleigh-Ritz variational method \cite{BransdenJoachain2003}. In the calculations presented here, we have taken \targetwave to be the target wavefunction described in detail in another of our previous calculations \cite{CooperArmourPlummer2008}, accounting for $96.8 \%$ of the correlation energy \cite{BransdenJoachain2003} of \hmol. The function, 

\begin{eqnarray}\nonumber
 \chi_{0} = &&\frac{N}{\lambda_{3} - 1}\cos\left[c\left( \lambda_{3}-1\right) \right]\\ 
  \label{eq:Chi}&\times&\lbrace 1-\exp\left[-\gamma\left( \lambda_{3}-1\right)\right]\rbrace\exp\left[-\gamma\left( \lambda_{3}-1\right)\right],
\end{eqnarray}
is the same as has been used in our earlier calculations \cite{Cooper2007,CooperArmourPlummer2008} and was introduced first by Massey and Ridley \cite{MasseyRidley1956}. The remaining short-range correlation functions, $\Omega=\{\chi_{1},\dots,\chi_{M}\}$, allow for the description of direct electron-positron and electron-electron interactions. Here, we have used the same set of $M=279$ correlation functions described in detail in equations (5-8) of \cite{CooperArmourPlummer2008}. The general form of each function, $\chi_{i}$, is

\begin{equation}\label{eq:srcf}
 \chi_{i}=f_{i}\left(\mathbf{r}_{1},\mathbf{r}_{2},\mathbf{r}_{3}\right)\exp\left[-\beta\left(\lambda_{1}+\lambda_{2}\right)-\alpha\lambda_{3}\right] \quad \left(1\leq i \leq M\right),
\end{equation}
where each $f_{i}\left(\mathbf{r}_{1},\mathbf{r}_{2},\mathbf{r}_{3}\right)$ is symmetric in the coordinates of the electrons. They are a mixture of separable correlation functions and Hylleraas-type functions \cite{Hylleraas1929} containing the electron-positron distance as a linear factor. As discussed previously \cite{CooperArmourPlummer2008}, the Hylleraas-type functions in particular allow for high accuracy of results away from anomalous singularities. Unless otherwise noted, we have here chosen values of $\alpha=0.6$ and $\beta=1.0$, rather than the values of $\alpha=0.3$ and $\beta=0.7$ used earlier \cite{CooperArmourPlummer2008}. This choice of nonlinear parameters highlights the interesting aspects of Schwartz-type anomalies more clearly.

In our application of the Kohn variational principle \cite{Cooper2007,CharltonHumberston2005}, the functional

\begin{equation}\label{eq:principle}
 \mathcal{J}\left[\Psi_{\mathrm{t}}\right]=\tan\left(\eta_{\mathrm{v}}-\tau+c\right)=a_{\mathrm{t}}-\frac{2}{\pi N^2 R^2 k}\langle\Psi_{\mathrm{t}},\Psi_{\mathrm{t}}\rangle
\end{equation}
is made stationary with respect to variations in $a_{\mathrm{t}}$ and $\{p_{0},\dots,p_{M}\}$. Here, we have denoted by $\langle\Psi_{\mathrm{t}},\Psi_{\mathrm{t}}\rangle$ the integral, $\langle \Psi_{\mathrm{t}}\vert\left(\hat{H}-E\right)\vert\Psi_{\mathrm{t}}\rangle$, where $\hat{H}$ is the nonrelativistic Hamiltonian for the scattering system and $E$ is the sum of the positron kinetic energy and the ground state energy expectation value of \targetwave. The integral is evaluated over the configuration space of the positron and the two electrons. We will, henceforth, use this notation more generally to denote by $\langle X,Y\rangle$ integrals of the form $\langle X\vert\left(\hat{H}-E\right)\vert Y\rangle$.

The stationary principle imposed upon (\ref{eq:principle}) leads to the linear system of equations

\begin{equation}\label{eq:KohnEq}
 Ax=-b,
\end{equation}
where

\begin{eqnarray}\label{eq:matA}
A&=&\left[ \begin{array}{cccc} 
\langle\bar{C}\psi_{\mathrm{G}},\bar{C}\psi_{\mathrm{G}}\rangle & \langle\bar{C}\psi_{\mathrm{G}},\chi_{0}\psi_{\mathrm{G}}\rangle &
\cdots& 
\langle\bar{C}\psi_{\mathrm{G}},\chi_{M}\rangle \\ 
\langle\chi_{0}\psi_{\mathrm{G}},\bar{C}\psi_{\mathrm{G}}\rangle & \langle\chi_{0}\psi_{\mathrm{G}},\chi_{0}\psi_{\mathrm{G}}\rangle &
\cdots &
\langle\chi_{0}\psi_{\mathrm{G}},\chi_{M}\rangle \\
\vdots & \vdots & \ddots & \vdots \\
 \langle\chi_{M},\bar{C}\psi_{\mathrm{G}}\rangle & \langle\chi_{M},\chi_{0}\psi_{\mathrm{G}}\rangle &
\cdots & 
\langle\chi_{M},\chi_{M}\rangle
\end{array}\right],\\
 \label{eq:vecB}b&=&\left[\begin{array}{c}
          \langle\bar{C}\psi_{\mathrm{G}},\bar{S}\psi_{\mathrm{G}}\rangle \\
	 \langle\chi_{0}\psi_{\mathrm{G}},\bar{S}\psi_{\mathrm{G}}\rangle \\
	 \vdots \\
	 \langle\chi_{M},\bar{S}\psi_{\mathrm{G}}\rangle \\
         \end{array}\right],\\
x&=&\left[\begin{array}{c}
          a_{t}\\
	  p_{0}\\
	  \vdots\\
 	  p_{M}
          \end{array}\right].
\end{eqnarray}
Solving (\ref{eq:KohnEq}) determines the values of $a_{\mathrm{t}}$ and $\{p_{0},\dots,p_{M}\}$, allowing $\langle\Psi_{\mathrm{t}},\Psi_{\mathrm{t}}\rangle$ and, hence, \ps to be calculated via (\ref{eq:principle}). However, as has been discussed extensively (see, for example, \cite{Cooper2007,Armour1987,Armour1988,Armour2008}), the particular form of the functions, $f_{i}\left(\mathbf{r}_{1},\mathbf{r}_{2},\mathbf{r}_{3}\right)$, used in our calculations does not, in general, permit analytic evaluation of the integrals comprising the matrix elements of $A$ and $b$. Sophisticated methods to determine these integrals numerically have been developed \cite{Armour1987,Armour1988,Armour2008}. However, the numerical approaches can give only accurate approximations to the exact values of the integrals, so that small errors in determining the elements of $A$ and $b$ are unavoidable.

\subsection{Singularities}\label{ss:singularities}

Singularities in our generalized Kohn calculations arise from zeros of \deter, the determinant of (\ref{eq:matA}). Under these circumstances, the linear system (\ref{eq:KohnEq}) has no unique solution. Close to these singularities, it is well known \cite{Schwartz1961,Schwartz1961b,Nesbet1968,Brownstein1968,Shimamura1971} that values of \ps obtained by solving (\ref{eq:KohnEq}) can be anomalous; small errors in the elements of $A$ or $b$ can correspond to large errors in the solution, $x$, particularly when $A$ is close to singularity in a sense that can be defined formally in terms of the condition number \cite{Higham2002,ArmourHumberston1991} of $A$. A more detailed discussion of the condition number will be given in section \ref{ss:persistent}. 

It is appropriate at this point to define a convention that we will adopt in our discussion of singularities in the generalized Kohn method. The type of spurious singularities mentioned by Schwartz \cite{Schwartz1961,Schwartz1961b} here correspond to zeros in \deter for the particular case when $\tau=0$. We will, however, find it convenient to label as Schwartz singularities those zeros of \deter occurring at any $\tau_{\mathrm{s}}\in\left[0,\pi\right)$ which give rise to anomalous behaviour in the calculations of $\eta_{\mathrm{v}}\left(\tau\right)$ when $\tau$ is near $\tau_{\mathrm{s}}$. This is an important clarification for the following reason: we claim that, because of our inclusion of $\tau$ in \trialwave, there exist zeros of \deter which are not spurious and which do not correspond to anomalous behaviour in the values of \ps. We will refer to such singularities as anomaly-free singularities.

To understand how anomaly-free singularities might arise, it is helpful to consider the component, $\Psi_{0}$, of the exact scattering wave function, $\Psi$, corresponding to the lowest partial wave. $\Psi_{0}$ can be expanded as 

\begin{equation}\label{eq:exact}
\Psi_{0}=\left(\bar{S}+a\bar{C}\right)\psi+\sum_{i=1}^{\infty}p_{i}\zeta_{i},
\end{equation}
where $\psi$ is the exact ground state target wavefunction and the complete set of correlation functions, $\{\zeta_{i}\}$, describes exactly the leptonic interactions at short-range. As noted by Takatsuka and Fueno \cite{Takatsuka1979} in their Kohn calculations for single channel scattering, the exact phase shift, $\eta_{0}$, determined by $\Psi_{0}$ is independent of the choice of $\tau$ in (\ref{eq:exact}). As a result, there is precisely one value, $\tau_{0}\in\left[0,\pi\right)$, at each positron energy such that

\begin{equation}
 \eta_{0}-\tau_{0}+c=\pm n\frac{\pi}{2},
\end{equation}
for some odd value of $n>0$, where either $+n$ or $-n$ is chosen to keep $\eta_{0}\in\left(-\pi/2,\pi/{2}\right]$. The value of $\cot\left(\eta_{0}-\tau+c\right)$ will then pass continuously through zero as $\tau$ passes through $\tau_{0}$. 

Returning to the trial wavefunction (\ref{eq:trialwave}), for nonsingular $A$ it can be shown using (\ref{eq:principle}) and (\ref{eq:KohnEq}) that

\begin{equation}
\cot\left(\eta_{\mathrm{v}}-\tau+c\right)=\frac{\pi N^2 R^2 k}{2}\left(\frac{\det\left(A\right)}{\hat{x}.b-\det\left(A\right)\langle\bar{S}\psi_{\mathrm{G}},\bar{S}\psi_{\mathrm{G}}\rangle}\right),
\end{equation}
where we have defined

\begin{equation}
 \hat{x}=\left[\mathrm{adj}\left(A\right)\right] b,
\end{equation}
$\mathrm{adj}\left(A\right)$ being the adjugate matrix of (\ref{eq:matA}). We note that $\mathrm{adj}\left(A\right)$ exists even if $A$ is singular. In section \ref{ss:pscalcs} we will provide numerical evidence that the limit of $\cot\left(\eta_{\mathrm{v}}-\tau+c\right)$ as $\det\left(A\right)\rightarrow0$ exists and is equal to zero. We can, therefore, assert a correspondence between zeros of $\cot\left(\eta_{\mathrm{v}}-\tau+c\right)$ and zeros of \deter.

Suppose that, at each $k$, there are $m_{k}$ values of $\tau$ making $A$ singular, which we will denote by $\Upsilon_{k}=\{\tau_{\mathrm{s}}^{\left(k,1\right)},\tau_{\mathrm{s}}^{\left(k,2\right)},\dots,\tau_{\mathrm{s}}^{\left(k,m_{k}\right)}\}$. If our previous assertion is correct, each element of $\Upsilon_{k}$ will correspond to a zero of $\cot\left(\eta_{\mathrm{v}}-\tau+c\right)$. We can reasonably claim that if \trialwave is, in some sense, sufficiently accurate, precisely one of these zeros will correspond to the zero of $\cot\left(\eta_{0}-\tau+c\right)$ necessarily found at $\tau_{0}$ for $\Psi_{0}$. Assuming that this is the case, we will denote by $\hat{\tau}_{\mathrm{s}}$ the element of $\Upsilon_{k}$ corresponding to $\tau_{0}$. Values of \ps determined in the generalized Kohn method should then vary slowly and smoothly with $\tau$ as it passes through $\hat{\tau}_{\mathrm{s}}$. Indeed, as $\tau\rightarrow\hat{\tau}_{\mathrm{s}}$ from either side, we would expect the values of \ps calculated by solving (\ref{eq:KohnEq}) to converge to the value, \psh, determined directly from

\begin{equation}\label{eq:afm}
 \hat{\eta}_{\mathrm{v}}-\hat{\tau}_{\mathrm{s}}+c = \pm n\frac{\pi}{2},
\end{equation}
where $n$ is again chosen so that $\hat{\eta}_{\mathrm{v}}\in\left(-\pi/2,\pi/2\right]$.

In the following section we will present results of generalized Kohn calculations exhibiting anomalous behaviour due to Schwartz singularities and, further, demonstrate empirically that the anomaly-free singularities do exist and that values of $\hat{\tau}_\mathrm{s}$ can be found. At each $k$, choosing $\tau=\hat{\tau}_\mathrm{s}\left(k\right)$ then defines an optimization of $\tau$ that will be seen to avoid anomalies in \ps due to Schwartz singularities.

\section{Results}
\subsection{Calculations of phase shift}\label{ss:pscalcs}

\begin{figure}
 \centering
 \includegraphics{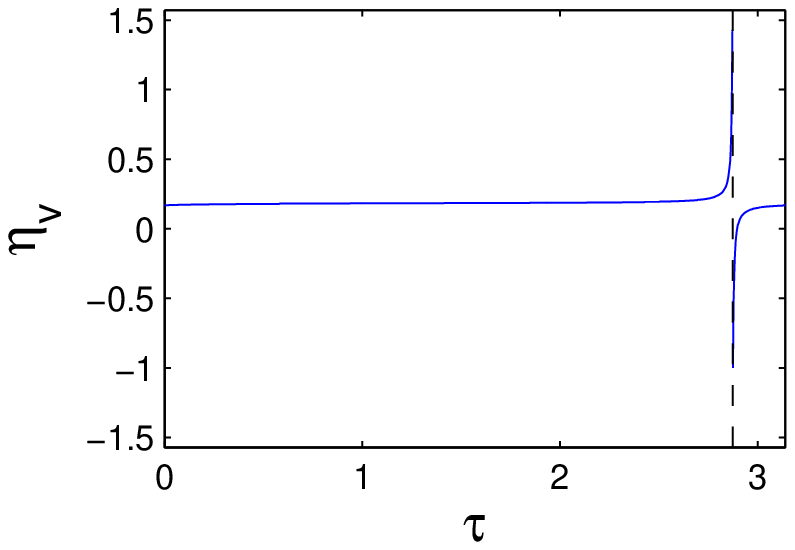}
 \caption{Values of $\eta_{\mathrm{v}}\left(\tau\right)$ at $k=0.2$.}
 \label{fig:ps2D}
\end{figure}

In our generalized Kohn calculations, we have obtained values of $\eta_{\mathrm{v}}\in\left(-\pi/2,\pi/2\right]$ using (\ref{eq:principle}) and (\ref{eq:KohnEq}), for a range of positron momenta. Spurious singularities have been accounted for by performing calculations over $p$ different values of $\tau$ equidistant in the range $\tau\in\left[0,\pi\right)$. For the results presented here, we have taken $p=1001$. Calculations for a large number of $\tau$ values can be carried out with minimal additional computational effort, as it can be shown that the matrix elements of $A$ and $b$ for any $\tau$ are readily available from the elements of $A\left(\tau=0\right)$ and $b\left(\tau=0\right)$ via an orthogonal transformation. It is helpful to carry out calculations for large $p$ as it allows a detailed examination of the behaviour of \ps very close to Schwartz singularities to be made. Values of \ps over the $p$ values of $\tau$ are given in figure \ref{fig:ps2D} for $k=0.2$, corresponding to a positron energy of $0.54$ eV. Anomalous results due to a Schwartz singularity are clearly evident around $\tau\sim2.87$. We have indicated the value of $\tau$ giving rise to the singularity by a dashed line. Away from this value of $\tau$, the variation in \ps is small.

\begin{figure}
 \centering
 \includegraphics{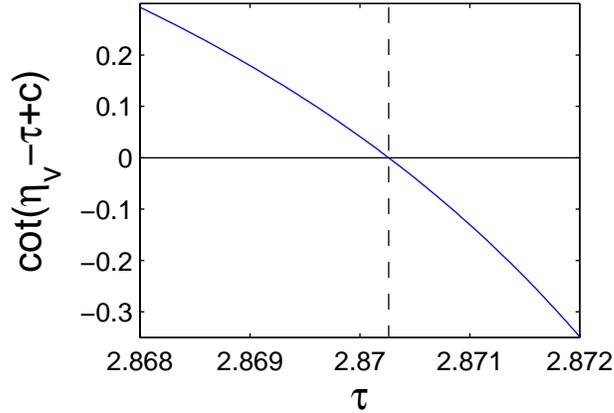}
 \caption{The behaviour of $\cot\left(\eta_{\mathrm{v}}-\tau+c\right)$ at $k=0.2$ for values of $\tau$ either side of a singularity.}
 \label{fig:cot}
\end{figure}

In figure \ref{fig:cot} we have studied the behaviour of $\cot\left(\eta_{\mathrm{v}}-\tau+c\right)$ close to the singularity at $\tau\sim2.87$, by calculating $\cot\left(\eta_{\mathrm{v}}-\tau+c\right)$ for $101$ values of $\tau$ equidistant in the range $\tau\in\left[2.868,2.872\right]$. We have again indicated the position of the singularity in this figure by a dashed line. The results shown in the figure suggest that $\cot\left(\eta_{\mathrm{v}}-\tau+c\right)$ converges smoothly to zero as $\tau\rightarrow\tau_{\mathrm{s}}$ from either side, supporting the assertion made in section \ref{ss:singularities} regarding the correspondence of the zeros in $\cot\left(\eta_{\mathrm{v}}-\tau+c\right)$ and \deter. We have found that behaviour of the type shown in figure \ref{fig:cot} is a general feature of the calculation.

Using (\ref{eq:tautrans}) and (\ref{eq:matA}), it is straightforward to show that

\begin{equation}\label{eq:detan}
 \det\left(A\right) = \mathcal{A}\left(k\right)\sin^{2}\left(\tau\right)+\mathcal{B}\left(k\right)\sin\left(\tau\right)\cos\left(\tau\right)+\mathcal{C}\left(k\right)\cos^{2}\left(\tau\right),
\end{equation}
where $\mathcal{A}\left(k\right)$, $\mathcal{B}\left(k\right)$ and $\mathcal{C}\left(k\right)$ are constants with respect to variations in $\tau$. For a given positron momentum, the constants, $\mathcal{A}$, $\mathcal{B}$ and $\mathcal{C}$, can be determined by calculating \deter directly from $(\ref{eq:matA})$ at particular values of $\tau$. Strictly speaking, in our calculations we have evaluated $\det\left(\tilde{A}\right)$, $\tilde{A}$ being the approximation to $A$ whose elements have been determined using numerical integration. We will assume that the values of $\mathcal{A}$, $\mathcal{B}$ and $\mathcal{C}$ are not unduly sensitive to small changes in the elements of $A$ and henceforth take $\det\left(A\right)$ and $\det\left(\tilde{A}\right)$ to be essentially equivalent. 

\begin{figure}
 \centering
 \includegraphics{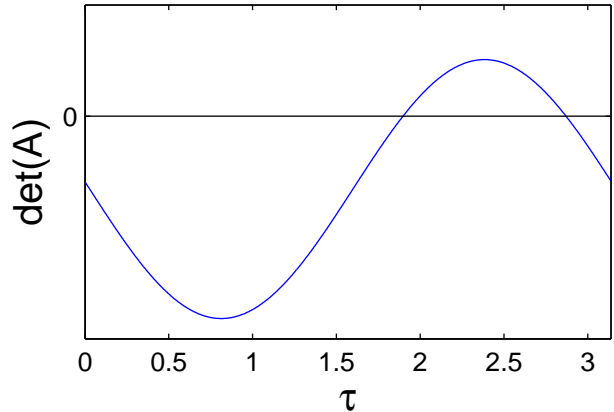}
 \caption{Values of $\det\left(A\right)$ at $k=0.2$ for $0\leq\tau<\pi$.}
 \label{fig:det2D}
\end{figure}

At each $k$, provided that $\mathcal{A}\neq0$, the values, $\tau_{\mathrm{s}}$, making $A$ singular can be found by solving the quadratic equation in $\tan\left(\tau_{\mathrm{s}}\right)$,

\begin{equation}\label{eq:qeq}
 \mathcal{A}\tan^{2}\left(\tau_{\mathrm{s}}\right)+\mathcal{B}\tan\left(\tau_{\mathrm{s}}\right)+\mathcal{C}=0.
\end{equation}
If only $\tau$ is varied, and unless $\mathcal{A}=\mathcal{B}=\mathcal{C}=0$, there will be no more than two zeros of \deter in the range $\tau\in\left[0,\pi\right)$. Figure \ref{fig:det2D} shows \deter as a function of $\tau$ at $k=0.2$. The scale on the vertical axis is unimportant, since the value of \deter at each $k$ can be made arbitrarily large or small by a choice of the normalization constant, $N$. The result of interest in the figure is that it indicates two values of $\tau$ at which $A$ is singular. 
The anomalous behaviour in figure \ref{fig:ps2D} corresponds directly to the singularity observed at $\tau\sim2.87$ in figure \ref{fig:det2D}. However, there are no anomalies in figure \ref{fig:ps2D} corresponding to the singularity at $\tau\sim1.90$ in figure \ref{fig:det2D}, suggesting that this singularity is of the anomaly-free type described in the previous section. 

\begin{figure}
 \centering
 \includegraphics{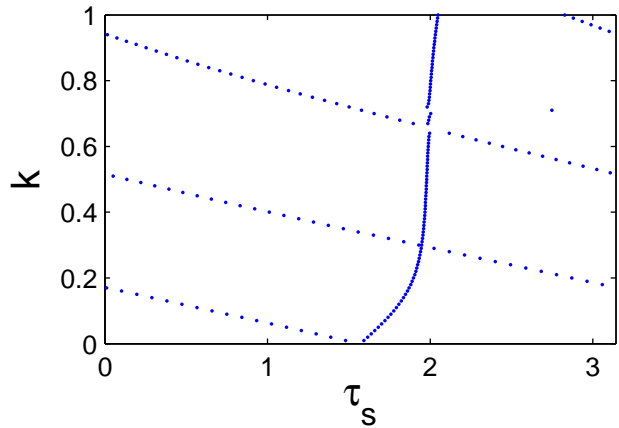}
 \caption{The zeros of $\det\left(A\right)$ for $0.01\leq k\leq 1$.}
 \label{fig:sMap}
\end{figure}

We have examined this phenomenon at other values of the positron momentum. Figure \ref{fig:sMap} indicates the roots of (\ref{eq:qeq}) for $100$ different positron momenta equidistant in the range $0.01\leq k \leq 1$, corresponding to a positron energy range from $1.36$ meV to $13.6$ eV. For the majority of positron momenta considered here, $\mathcal{A}$, $\mathcal{B}$ and $\mathcal{C}$ are such that there are two values of $\tau_{\mathrm{s}}$ at each $k$, the exceptions being $k=0.65$ and $k=0.66$, for which we have found no real-valued solutions of (\ref{eq:qeq}). It is apparent from figure \ref{fig:sMap} that the roots of (\ref{eq:qeq}) lie in two families of curves. The first family spans the entire range, $\tau_{\mathrm{s}}\in\left[0,\pi\right)$, for $0.01\leq k \leq1$. The second family is confined to values of $\tau_{\mathrm{s}}$ in the range $\tau_{\mathrm{s}}\in\left[1.5,2.1\right]$ for all positron momenta considered here. For almost every $k$ where real roots exist, there is precisely one root corresponding to each family. The exception is the result at $k=0.71$, where there is an irregularity in the otherwise smooth behaviour of $\tau_{\mathrm{s}}$ over $k$. We will discuss this phenomenon in more detail in section \ref{ss:persistent}. 

\begin{figure}
 \centering
 \includegraphics{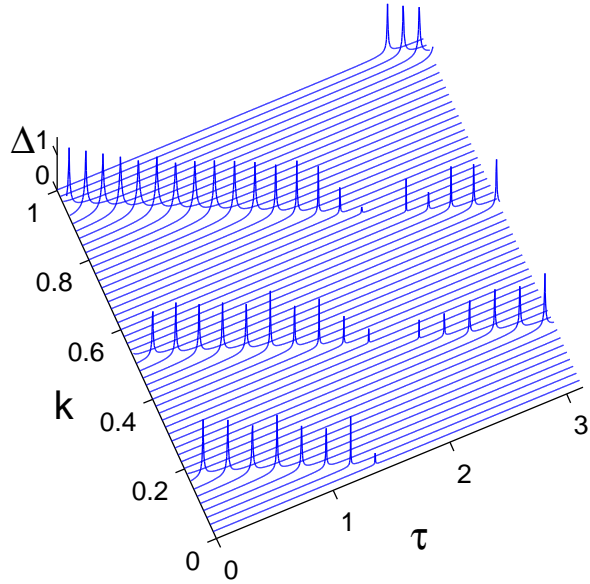}
 \caption{Values of $\Delta\left(k,\tau\right)$ for $0.02\leq k \leq1$.}
 \label{fig:psMap}
\end{figure}

To illustrate anomalies in \ps at different values of $k$, it is convenient to define the function,

\begin{equation}\label{eq:Delta1}
 \Delta\left(k,\tau\right)=\vert\eta_{\mathrm{v}}\left(k,\tau\right)-\langle{\eta}_{\mathrm{v}}\rangle\left(k\right)\vert,
\end{equation}
where, at each $k$, $\langle{\eta}_{\mathrm{v}}\rangle$ is the median value of $\eta_{\mathrm{v}}\left(\tau\right)$ evaluated over the $p$ values of $\tau$. $\Delta\left(k,\tau\right)$ measures the degree to which a given $\eta_{\mathrm{v}}\left(k,\tau\right)$ can be considered anomalous. The values of $\Delta\left(k,\tau\right)$ are shown in figure \ref{fig:psMap}. For clarity, we have included results only for $50$ values of $k$ equidistant in the range $0.02\leq k \leq1$, rather than the $100$ values used for figure \ref{fig:sMap}. The omission of the results for $k=0.71$ in this figure also allows us to delay until section \ref{ss:persistent} the discussion of the atypical singularity observed at this value of $k$ in figure \ref{fig:sMap}.  

\begin{figure}
 \centering
 \includegraphics{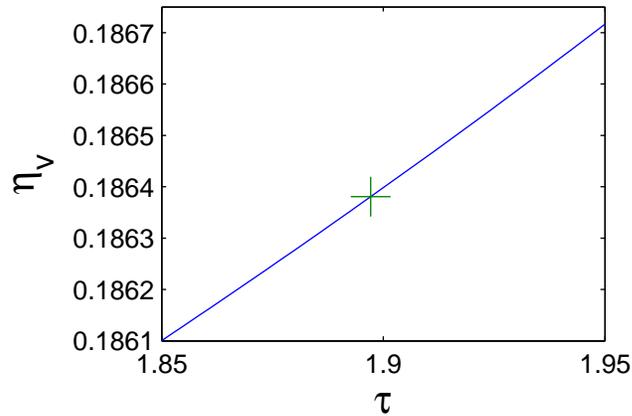}
 \caption{Convergence of [\textcolor{blue}{---}]$\eta_{\mathrm{v}}\left(\tau\right)$ to [\textcolor{myGreen}{+}]$\hat{\eta}_{\mathrm{v}}$ as $\tau\rightarrow\hat{\tau}_{\mathrm{s}}$ at $k=0.2$.}
 \label{fig:convergence}
\end{figure}

It is clear from figure \ref{fig:psMap} that anomalies are observed corresponding to only the first of the two families of curves identified from figure \ref{fig:sMap}. This is strong evidence that the curve for which no anomalies are observed comprises legitimately occurring singularities. It is interesting to note that the size of the anomalies due to the Schwartz singularities become noticeably smaller in figure \ref{fig:psMap} as they tend to coincide with the apparently anomaly-free singularities. Denoting by $\tau_{\mathrm{s}}^{\left(1\right)}\left(k\right)$ the values of $\tau$ describing the anomaly-free curve, at each $k$ we expect that, as $\tau\rightarrow\tau_{\mathrm{s}}^{\left(1\right)}$, the values of \ps calculated over $\tau$ will converge smoothly to the value, $\hat{\eta}_{\mathrm{v}}$, determined directly from (\ref{eq:afm}) and taking $\hat{\tau}_{\mathrm{s}}=\tau_{\mathrm{s}}^{\left(1\right)}$. We have found in our calculations that this is indeed the case, and an example is shown in figure \ref{fig:convergence} for $k=0.2$. 

\begin{figure}
 \centering
 \includegraphics{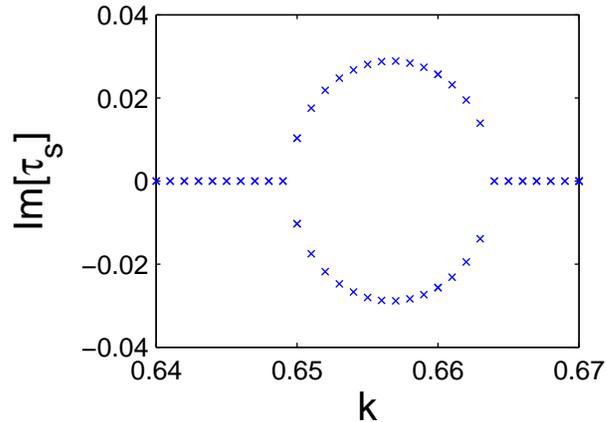}
 \caption{Values of $\Im\left[\tau_{\mathrm{s}}\right]$ for $0.64 \leq k \leq 0.67$.}
 \label{fig:imroot}
\end{figure}

We can reasonably conclude that we have encountered anomaly-free singularities. Before we can develop an optimization for $\tau$ based on these singularities, however, there are two outstanding issues to be addressed. Firstly, we have already noted that no real-valued solutions of (\ref{eq:qeq}) were found at $k=0.65$ and $k=0.66$. This is potentially problematic as, for a sufficiently accurate trial function, we should expect at least one real root of (\ref{eq:qeq}) at each $k$, corresponding to an anomaly-free singularity. However, inspection of figure \ref{fig:sMap} shows that the solutions of (\ref{eq:qeq}) are close together in the regions either side of $k=0.65$ and $k=0.66$. Were the two roots to coincide at some $k$, then $\mathcal{B}^2=4\mathcal{A}\mathcal{C}$. Near a point of coincidence, $\mathcal{B}^2\sim4\mathcal{A}\mathcal{C}$ and small errors in the values of $\mathcal{A}$, $\mathcal{B}$ and $\mathcal{C}$ could erroneously give rise to $\left(\mathcal{B}^2-4\mathcal{A}\mathcal{C}\right)<0$. The values of $\tau$ solving (\ref{eq:qeq}) at $k=0.65$ and $k=0.66$ were found to be, respectively, $\tau_{\mathrm{s}}\sim2.02\pm0.01\rmi$ and $\tau_{\mathrm{s}}\sim1.98\pm0.03\rmi$. In both cases, the fact that $\Im\left[\tau_{\mathrm{s}}\right]\ll\Re\left[\tau_{\mathrm{s}}\right]$ suggests that singularities do genuinely exist for $\tau\in\mathbb{R}$ at these values of $k$, but small errors in our calculations of $\mathcal{A}$, $\mathcal{B}$ and $\mathcal{C}$ due to inexact numerical integration have prevented us from finding them. Having investigated this problem in more detail, in figure \ref{fig:imroot} we show the calculated values of $\Im\left[\tau_{\mathrm{s}}\right]$ for $31$ values of $k$ equidistant in the range $0.64 \leq k \leq 0.67$. There is a clearly defined region of $k$ where no real roots of (\ref{eq:qeq}) have been found. The smoothness of $\Im\left[\tau_{\mathrm{s}}\right]$ over $k$ in this region does not necessarily preclude the notion that the failure to find real-valued solutions is due to small numerical errors in our calculations. It is conceivable that inaccuracies in the calculated values of $\mathcal{A}$, $\mathcal{B}$ and $\mathcal{C}$ could also arise from systematic errors in the algorithm \cite{NAG1} used to calculate the determinants. Nevertheless, the results illustrated in figure \ref{fig:imroot} are interesting; their exact origin may be speculated upon and will remain a subject of our ongoing investigations.

The second difficulty concerns the choice of $\hat{\tau_{\mathrm{s}}}$ from the two available solutions of (\ref{eq:qeq}). A method is needed for identifying at each $k$ the root of (\ref{eq:qeq}) corresponding to a legitimate singularity. This can easily be achieved by inspecting values of \ps at values of $\tau$ either side of each singularity, although this approach is not ideal as it requires solutions of (\ref{eq:KohnEq}) to be found. In practice, at each $k$ it should be possible to determine by inspection which of the two phase shifts is anomaly-free by examining corresponding results at singularities for nearby values of $k$. For example, figure \ref{fig:sMap} clearly shows that only one curve in the $\left(\tau,k\right)$ plane corresponds to a physically acceptable variation of phase shift over $k$. 

\begin{figure}
 \centering
 \includegraphics{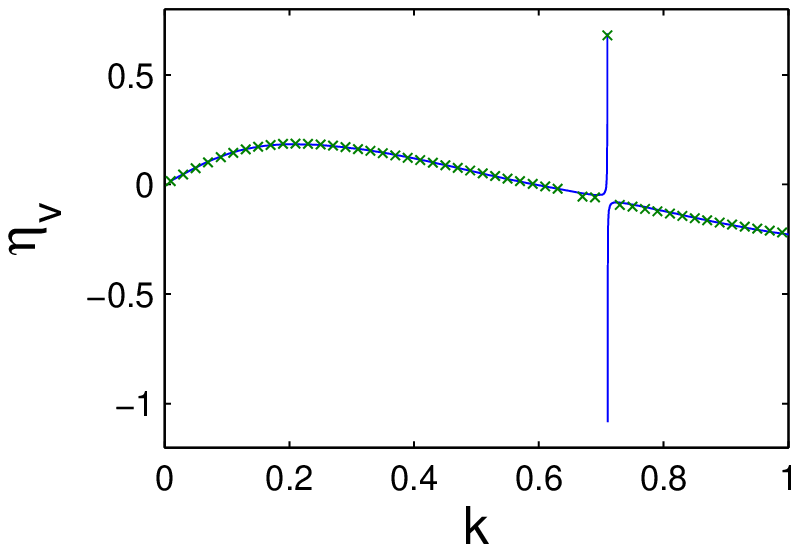}
 \caption{A comparison of optimization schemes for $\tau$, [\textcolor{blue}{\full}] $\langle{\eta}_{\mathrm{v}}\rangle\left(k\right)$ and [\textcolor{myGreen}{$\times$}] $\hat{\eta}_{\mathrm{v}}\left(k\right)$.}
 \label{fig:comparison}
\end{figure}

With these considerations in mind, we claim that choosing $\tau=\hat{\tau}_\mathrm{s}$ at each $k$ defines a consistent optimization that can be used to avoid anomalies due to Schwartz singularities appearing at other values of $\tau$. To evaluate the success of this approach, we have found it helpful to consider an alternative optimization of $\tau$. For the calculations of \ps carried out with $p=1001$, choosing the value of $\tau$ at each $k$ giving rise to the median phase shift, $\langle\eta_{\mathrm{v}}\rangle$, should also mitigate anomalous behaviour. In figure \ref{fig:comparison} we have compared results for $\langle{\eta}_{\mathrm{v}}\rangle\left(k\right)$ and $\hat{\eta}_{\mathrm{v}}\left(k\right)$ for momenta in the range $0.01\leq k \leq1$. Here, of the two candidates for $\hat{\eta}_{\mathrm{v}}$ corresponding to the two singularities, at each $k$ we have chosen the one whose absolute value is closest to $\vert\langle{\eta}_{\mathrm{v}}\rangle\vert$. For clarity, we have included in the figure values of $\hat{\eta}_{\mathrm{v}}\left(k\right)$ for only $50$ values of $k$ equidistant in the range $0.01 \leq k \leq 0.99$.

Both optimization schemes successfully avoid anomalous behaviour at most values of $k$ and there is good agreement between the two sets of results at all momenta. However, the intriguing feature of the figure is the anomaly appearing in both sets of results at $k\sim0.71$. For $\langle{\eta}_{\mathrm{v}}\rangle\left(k\right)$, we have shown this anomaly in greater detail by including in the figure results of a further $100$ Kohn calculations for momenta equidistant in the range $0.7\leq k\leq0.72$, although it is practical to consider henceforth only the anomalous behaviour occurring precisely at $k=0.71$. We believe the anomaly shown in figure \ref{fig:comparison} is of a different type to those shown in figure \ref{fig:psMap}, which are due to singularities found by varying only $\tau$ at a given $k$. In the following section we will examine the circumstances under which persistent anomalies of the kind shown in figure \ref{fig:comparison} could appear, before going on to discuss methods designed to avoid them.

\subsection{Persistent anomalous behaviour}\label{ss:persistent}

It is often claimed that anomalous results observed in the region of singularities arise from $A$ having a determinant close to zero. Statements of this kind can be misleading, as the determinant of any nonsingular $A$ can be made arbitrarily close to zero by an appropriate scalar multiplication, without altering the sensitivity of the solution, $x$, to small errors in the elements of $A$ or $b$. A better measure for identifying regions where anomalies may occur is the condition number, \cond, defined for nonsingular $A$. The condition number is independent of the normalization constant, $N$, and for the linear system (\ref{eq:KohnEq}) is defined as the maximum ratio of the relative error in $x$ and the relative error in $b$. Formally, it can be shown that
\begin{equation}\label{eq:condition}
 \kappa\left(A\right)=\parallel A\parallel\parallel A^{-1}\parallel,
\end{equation}
with respect to some matrix norm, $\parallel A \parallel$ \cite{Higham2002}. The value of \cond is dependent upon the choice of norm. In our calculations, we have considered the matrix 1-norm \cite{Higham2002} of $A$, 

\begin{equation}\label{eq:onenorm}
\parallel A \parallel_{1}=\max_{1\leq j\leq\left(M+2\right)}\sum_{i=1}^{M+2}\vert a_{ij}\vert,
\end{equation} 
where $a_{ij}$ is the element in the $i^{\mathrm{th}}$ row and $j^{\mathrm{th}}$ column of $A$. In what follows, the particular choice of the 1-norm in our calculations will implicitly be assumed.

A matrix with a large condition number is said to be ill-conditioned, and the solution of the corresponding linear system may not be reliable if the elements of $A$ and $b$ are not known exactly. For any invertible $A$, the condition number can be used to formalize the definition of closeness to singularity in the following way. If $\Delta A$ is defined to be any matrix such that $A+\Delta A$ is singular, then the relative distance to singularity, \dist, for $A$, is defined \cite{Higham2002} to be  

\begin{equation}\label{eq:distance}
 \Lambda\left(A\right)=\min\left(\frac{\parallel \Delta A\parallel}{\parallel A\parallel}:\quad \det\left(A+\Delta A\right)=0\right).
\end{equation}
This definition holds for any consistent norm. Further, if \dist and \cond are evaluated using the same choice of norm, it can be shown \cite{Higham2002} that 

\begin{equation}\label{eq:distcond}
 \Lambda\left(A\right) = \left[\kappa\left(A\right)\right]^{-1}.
\end{equation}

In section \ref{ss:pscalcs} we noted that (\ref{eq:qeq}) has, in general, no more than two zeros if only variations of $\tau\in\left[0,\pi\right)$ are considered. However, if $\mathcal{A}=\mathcal{B}=\mathcal{C}=0$ then \deter is identically zero independently of $\tau$ and no consistent value of phase shift can be calculated, either by solving (\ref{eq:KohnEq}) or directly from (\ref{eq:afm}). There is no obvious physical reason why this circumstance should arise at any $k\in\mathbb{R}$. However, it is conceivable that $\mathcal{A}$, $\mathcal{B}$ and $\mathcal{C}$ could coincidentally be close to zero, in some sense, over a narrow range of $k$. Small errors in the evaluation of $\mathcal{A}$, $\mathcal{B}$ and $\mathcal{C}$ could then give rise to both unreliable solutions of (\ref{eq:qeq}) and persistent anomalies in the calculation of \ps due to ill-conditioning in the Kohn equations (\ref{eq:KohnEq}). To see how the latter case arises, using (\ref{eq:detan}) we note that

\begin{eqnarray}
\mathcal{C}&=&\det\left[A\left(\tau=0\right)\right],\\
\mathcal{A}&=&\det\left[A\left(\tau=\frac{\pi}{2}\right)\right],\\
\mathcal{B}&=&2\det\left[A\left(\tau=\frac{\pi}{4}\right)\right]-\mathcal{A}-\mathcal{C},
\end{eqnarray}
so that the notion of $\mathcal{A}$, $\mathcal{B}$ and $\mathcal{C}$ being close to zero is immediately formalized in terms of \dist at $\tau=0$, $\tau=\pi/4$ and $\tau=\pi/2$.

\begin{figure}
 \centering
 \includegraphics{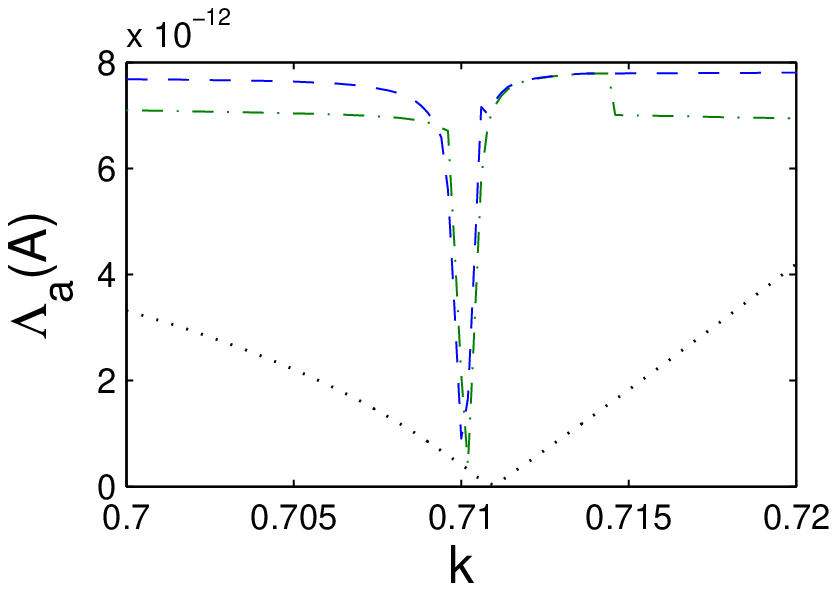}
 \caption{Values of \diste over $k$ for [\textcolor{blue}{\dashed}] $\tau=0$, [\textcolor{myGreen}{\chain}] $\tau=\frac{\pi}{4}$ and [{\dotted}] $\tau=\frac{\pi}{2}$.}
 \label{fig:ABC}
\end{figure}

In our calculations, we have used a numerical algorithm \cite{NAG2} to calculate \diste, an estimate of \dist. Values of \diste at $\tau=0$, $\tau=\pi/4$ and $\tau=\pi/2$ are shown in figure \ref{fig:ABC} for $100$ values of $k$ in the range $0.7 \leq k \leq 0.72$. Values of \diste for $\tau=0$ and $\tau=\pi/4$ are anomalously small at $k\sim0.71$, with the values for \diste at $\tau=\pi/2$ also passing through a clear minimum at $k\sim0.711$. Ordinarily, we would not expect ill-conditioning to occur over a very broad range of $\tau$ at a given $k$. At $k=0.71$, the small values of \diste at $\tau=0$, $\tau=\pi/4$ and $\tau=\pi/2$ therefore point to a manifestation of ill-conditioning which is unusually widespread in $\tau$. In fact, in our calculations we have failed to find any value of $\tau\in\left[0,\pi\right)$ such that $A$ is sufficiently well-conditioned to avoid anomalous results at $k=0.71$. We have also confirmed that $\mathcal{A}$, $\mathcal{B}$ and $\mathcal{C}$ all pass through zero at least once between $k=0.71$ and $k=0.7104$. 

We can conclude that both optimization schemes developed in section \ref{ss:pscalcs} successfully avoid anomalies due to Schwartz singularities whose existence depends only upon the choice of $\tau$. However, at certain values of $k$, $A$ can become close to singularities whose existence is independent of $\tau$, resulting in anomalies in the calculation of \ps that persist even after $\tau$ has been optimized. In attempting to address this problem, a number of methods are available. In principle, anomalous behaviour can be reduced dramatically by allowing the Kohn trial wavefunction to be complex-valued \cite{Schneider1988,McCurdy1987}. An alternative approach for avoiding anomalies persistent in $\tau$ is to make a small change in some other parameter of the trial wavefunction. We will explore both of these techniques in the following section.

\subsection{The complex Kohn method}

The complex Kohn method is an extension of the original variational approach in which the boundary conditions of the trial wavefunction are complex. It was originally believed \cite{Schneider1988,McCurdy1987} that this method was anomaly-free, although anomalies were subsequently reported by Lucchese \cite{Lucchese1989}. For our complex Kohn calculations on \ehmol scattering, we have used a trial wavefunction, $\Psi_{\mathrm{t}}'$, of the form

\begin{equation}\label{eq:complextrialwave}
\Psi_{\mathrm{t}}' = \left(\bar{S} + a_{\mathrm{t}}'\bar{T} + p_{0}'\chi_{0}\right)\psi_{\mathrm{G}} + \sum_{i=1}^{M} p_{i}'\chi_{i},
\end{equation}
where  

\begin{equation}\label{eq:hankel}
 \bar{T}=\bar{S}+\rmi\bar{C},
\end{equation}
the functions \targetwave and $\Omega=\{\chi_{1},\dots,\chi_{M}\}$ being the same as in (\ref{eq:trialwave}). The unknowns $a_{\mathrm{t}}'$ and $\{p_{0}',\dots,p_{M}'\}$ will not, in general, be real. Application of the variational principle to (\ref{eq:complextrialwave}) leads to a matrix equation analogous to (\ref{eq:KohnEq}),

\begin{equation}\label{eq:CKohnEq}
 A'x'=-b'
\end{equation}
where $A'$ and $b'$ are identical to $A$ and $b$, but for the function, $\bar{T}$, replacing $\bar{C}$ in (\ref{eq:matA}) and (\ref{eq:vecB}). The determinant, $\det\left(A'\right)$, conveniently reduces to

\begin{equation}
 \det\left(A'\right) = \mathcal{D}\left(k\right)e^{-2\rmi\tau},
\end{equation}
where 

\begin{equation}\label{eq:DABC}
 \mathcal{D}=\left(\mathcal{A}-\mathcal{C}\right)-\rmi\mathcal{B}
\end{equation}
is a complex constant with respect to variations in $\tau$. The values of $\det\left(A'\right)$ then describes a circle of radius $\vert \mathcal{D}\vert$ in the complex plane for variations of $\tau\in\left[0,\pi\right)$. Hence, singularities are obtained only if both the real and imaginary parts of $\mathcal{D}$ are zero; they can neither be located nor avoided by varying only $\tau$. We would therefore expect anomalous results due to singularities arising from the choice of $\tau$ at a given $k$ to be eliminated in the complex Kohn method. However, from (\ref{eq:DABC}) we would also expect $\mathcal{D}$ to be close to zero at $k=0.71$, in the same sense that $\mathcal{A}$, $\mathcal{B}$ and $\mathcal{C}$ have already been seen to be close to zero at this value of $k$. It is therefore likely that the anomalies already seen to occur due to relationships between $\mathcal{A}$, $\mathcal{B}$ and $\mathcal{C}$ rather than the choice of $\tau$, will persist even in the complex Kohn method.

\begin{figure}
 \centering
 \includegraphics{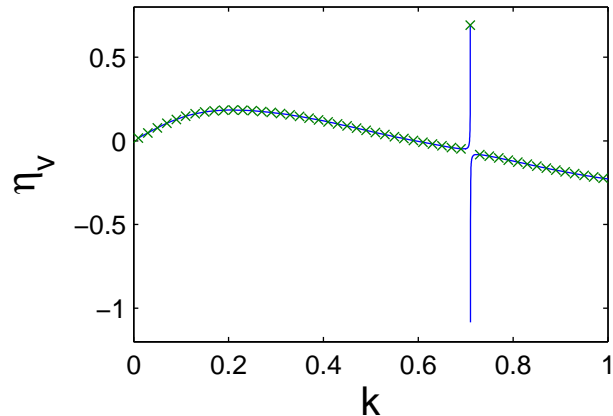}
 \caption{A comparison of [\textcolor{blue}{\full}] $\langle{\eta}_{\mathrm{v}}\rangle\left(k\right)$, determined using the trial function, \trialwave, with [\textcolor{myGreen}{$\times$}] $\eta_{\mathrm{v}}\left(k,\tau=0\right)$, evaluated using the complex Kohn method with the trial function, \ctrialwave.}
 \label{fig:methComp}
\end{figure}

We have obtained values of \ps using the trial function, \ctrialwave. We have found that the differences in the calculated values of \ps at different values of $\tau\in\left[0,\pi\right)$ are negligible, for all positron momenta considered here. Without loss of generality, we can regard the complex Kohn calculation as effectively independent of $\tau$ and choose $\tau=0$ for simplicity. In figure \ref{fig:methComp} we have compared results for $\eta_{\mathrm{v}}\left(k,\tau=0\right)$ obtained with the trial function, \ctrialwave, with the results for $\langle{\eta}_{\mathrm{v}}\rangle\left(k\right)$ obtained in section \ref{ss:pscalcs} with \trialwave. 

\begin{figure}
 \centering
 \includegraphics{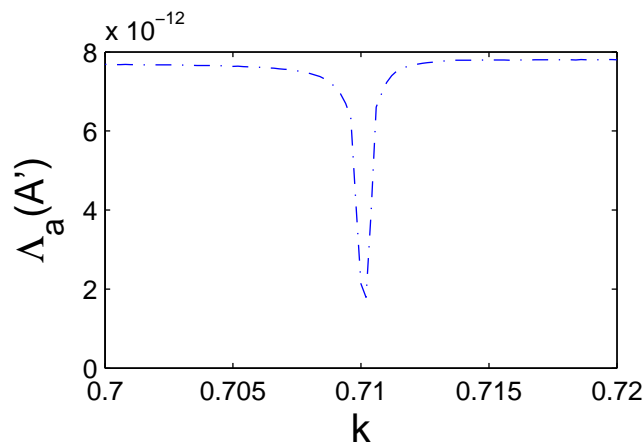}
 \caption{Values of $\Lambda_{\mathrm{a}}\left(A'\right)$ over $k$ for $\tau=0$.}
 \label{fig:D}
\end{figure}

The results of the two methods are essentially equivalent at all positron momenta, with the optimized results for \trialwave differing from the results for \ctrialwave by no more than $0.1 \%$ at each $k$. As expected, the complex Kohn method automatically avoids anomalies at most values of $k$ without the need for an optimization of $\tau$. Nevertheless, the use of the complex trial function has predictably failed to remove the persistent anomaly at $k=0.71$. In figure \ref{fig:D} we verify that $\Lambda_{\mathrm{a}}\left(A'\right)$ is anomalously small at $k\sim0.71$ for $\tau=0$. We have found that differences between the results shown in figure \ref{fig:D} and values of $\Lambda_{\mathrm{a}}\left(A'\right)$ calculated at other values of $\tau$ are negligible.

Having failed to find a systematic remedy for the persistent anomalous behaviour, we consider a more {\it ad hoc} approach. It should be possible to avoid any Schwartz-type anomaly by some variation of parameters in the trial wavefunction. We have found that variations in $\tau$ are not always successful, but other candidates exist. In our complex Kohn calculations, we have varied the values of $\alpha$ and $\beta$ in (\ref{eq:srcf}), fixing $\tau=0$. Recall that the values of the these parameters have so far remained fixed at $\alpha=0.6$ and $\beta=1.0$. We now consider the results of Kohn calculations carried out for $31$ different values of $\alpha$ in the range $\alpha\in\left[0.59,0.605\right]$ and $61$ different values of $\beta$ in the range $\beta\in\left[0.65,1.25\right]$.

\begin{figure}
 \centering
 \includegraphics{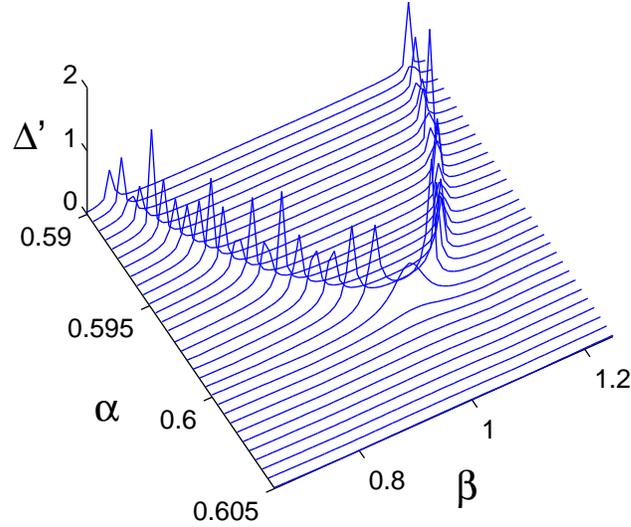}
 \caption{Values of $\Delta'\left(\alpha,\beta\right)$ at $k=0.71$.}
 \label{fig:pssSurf}
\end{figure}

To illustrate persistent anomalous behaviour, it is helpful to define a function analogous to (\ref{eq:Delta1}),

\begin{equation}\label{eq:Delta2}
\Delta'\left(\alpha,\beta\right)=\vert\eta_{\mathrm{v}}\left(\alpha,\beta\right)-\overline{{\eta}_{\mathrm{v}}}\left(\alpha\right)\vert,
\end{equation}
where, for each of the values of $\alpha$ considered, $\overline{{\eta}_{\mathrm{v}}}\left(\alpha\right)$ is the median value of \ps evaluated across the range of values of $\beta$. Values of $\Delta'\left(\alpha,\beta\right)$ are shown in figure \ref{fig:pssSurf}, from which it is clear that persistent anomalies appear distributed about a curve in the $\left(\alpha,\beta\right)$ plane. For values of $\alpha$ and $\beta$ away from this curve, the calculations are free of anomalies. Hence, a small change in the values of $\alpha$ or $\beta$ can indeed be shown to successfully avoid persistent anomalous behaviour.

\begin{figure}
 \centering
 \includegraphics{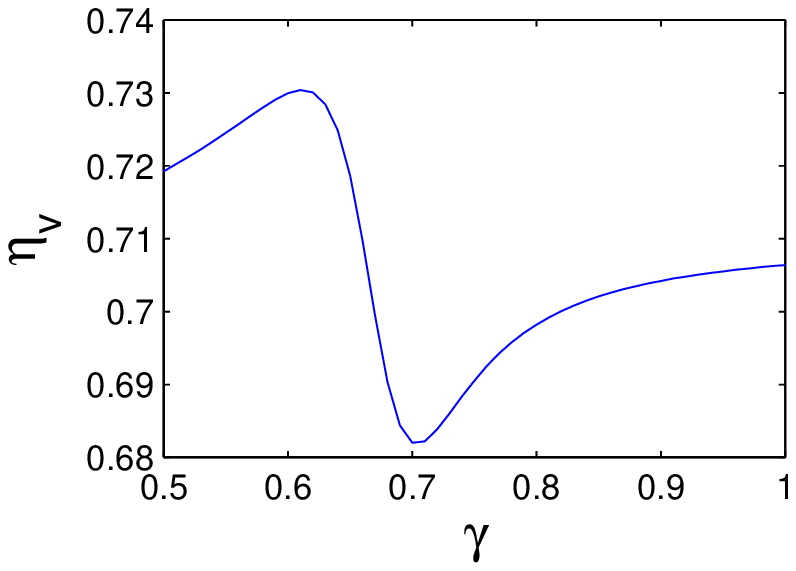}
 \caption{Values of $\eta_{\mathrm{v}}\left(\gamma\right)$ at $k=0.71$.}
 \label{fig:varyGamma}
\end{figure}

Finally, we consider briefly that the shielding parameter, $\gamma$, in (\ref{eq:CosOpenChannel}) and (\ref{eq:Chi}) might also be varied in an effort to avoid anomalous behaviour. Values of \ps at $k=0.71$, $\alpha=0.6$ and $\beta=1.0$ for $0.5\leq\gamma\leq1.0$ are shown in figure \ref{fig:varyGamma}. It is apparent that small changes in the value of $\gamma$ have relatively little effect on the persistent anomaly at $k=0.71$. This is not unexpected, being consistent with the findings of Lucchese \cite{Lucchese1989}, who investigated the effect of varying a parameter analogous to $\gamma$ in his model potential calculations. He noted that singularities due to the choice of $\gamma$ occurred when the values of $\gamma$ and $\alpha$ were not similar, most typically when $\gamma\ll\alpha$. With this in mind, and from inspection of figures \ref{fig:pssSurf} and \ref{fig:varyGamma}, we can conclude that the anomaly observed in figure \ref{fig:methComp} is due primarily to the choices of $\alpha$ and $\beta$ rather than the choice of $\gamma$.

\section{Concluding remarks}

We have carried out a thorough examination of singularities and related anomalous behaviour in generalized Kohn calculations for \ehmol scattering. We have argued that singularities do not always occur spuriously and that variational calculations of the scattering phase shift can be anomaly-free at these singularities. Subsequently, we have developed an optimization scheme for choosing a free parameter of the trial wavefunction allowing anomaly-free values of the phase shift to be determined without the need to solve the linear system of equations derived from the Kohn variational principle. This approach has been seen to be largely successful, giving phase shifts in close agreement with those determined by a conventional generalization of the Kohn method, as well as those obtained with the complex Kohn method. 

Persistent anomalies in both sets of calculations have been identified and attributed to singularities that cannot be avoided with any choice of the parameter, $\tau$. Further, we have found that our implementation of the complex Kohn method is susceptible to the same behaviour. We have demonstrated, however, that persistent anomalies can be avoided by small changes in the nonlinear parameters of the short-range correlation functions. Hence, by studying the behaviour of $\mathcal{A}$, $\mathcal{B}$ and $\mathcal{C}$ over $k$, we can predict the appearance of persistent anomalous behaviour quantitatively and avoid it by an appropriate change in $\alpha$ or $\beta$. 

\ack
We wish to thank John Humberston for valuable discussions. This work is supported by EPSRC (UK) grant EP/C548019/1.

\end{document}